\documentclass{PoS}
\usepackage{amsmath}
\usepackage{cite}
\newcommand{\C}[0]{{QCD}}
\newcommand{\E}[0]{{QED}}

\title{Dynamical QCD+QED simulation with staggered quarks}

\ShortTitle{Dynamical QCD+QED simulation with staggered quarks}

\author{\speaker{Ran Zhou}\\
	Department of Physics, Indiana University, Bloomington, IN 47405, USA\\
        Theoretical Physics Department, Fermi National Accelerator Laboratory, Batavia, 60510, USA\footnote{present address}\\
        E-mail: \email{zhouran@fnal.gov}}

\author{Steven Gottlieb\\
       Department of Physics, Indiana University, Bloomington, IN 47405, USA}

\author{MILC Collaboration}

\abstract{Electromagnetic effects play an important role in many
phenomena such as isospin-symmetry breaking in the 
hadron spectrum and the hadronic contributions to g-2.
We have generalized the MILC QCD code to include the
electromagnetic field.  In this work, we focus on 
simulations including charged sea quarks using the RHMC algorithm. 
We show details of the dynamical QCD+QED simulation algorithm with compact
QED. We analyze the code performance and results for 
hadron-spectrum observables.}

\FullConference{The 32nd International Symposium on Lattice Field Theory,\\
		23-28 June, 2014\\
		Columbia University New York, NY}

\begin{document}

\section{Introduction}
Electromagnetic corrections to phenomena dominated by
the strong interaction have received considerable attention in 
recent years.  For example, the electromagnetic effect on the hadron mass spectrum 
has been investigated extensively in lattice
simulations~\cite{Blum:2010ym, Horsley:2012fw, Basak:2012zx,  Basak:2013iw, Borsanyi:2013lga, deDivitiis:2013xla}. 
Although there has been much progress in this area, much of the research
relies upon the quenched QED approximation that neglects sea-quark 
charges.  There are some difficulties in estimating 
systematic errors associated with the quenched-QED approximation. Further, 
sea-quark electromagnetic effects are critical in some studies. For example, the 
sea-quark electromagnetic interaction contributes a term to the pseudoscalar meson mass, 
and one of the Low Energy Constants (LECs) in meson QCD+QED 
chiral perturbation theory (ChPT) is associated with this term.
To determine this LEC in a quenched-QED simulation, one has to use reweighing~\cite{Ishikawa:2012ix,Aoki:2012st}. 
In addition, the electromagnetic effect is also the dominant systematic in the current $m_u/m_d$ determination.
Similarly,  sea-quark  electromagnetic effects also play a 
role in the lattice-QCD muon anomalous magnetic moment calculation~\cite{Blum:2013qu}. 
The quark-disconnected contribution, for instance, is a significant source of the systematic 
error in the current calculations of the hadronic light-by-light term. 
The sea quarks are coupled to the electromagnetic field, and 
the sea-quark electromagnetic effect is not negligible.

Fully dynamical QCD+QED calculations include the $U(1)$ photon field
during configuration generation. This approach eliminates the unquantifiable quenching error. 
Recently, the QCDSF~\cite{Horsley:2013qka} and BMW~\cite{Borsanyi:2014jba} 
collaborations presented their dynamical QCD+QED studies of the 
meson and baryon masses based upon the non-compact QED formalism. In this work, we 
report on our progress developing a QCD+QED simulation code based on the compact
QED formalism.  

The structure of this paper is as follows. 
We describe the details of the numerical algorithm of the dynamical 
QCD+QED simulation in Sec.~\ref{sec:methodology}.  
We then show various test results related to configuration generation and 
the pseudoscalar meson spectrum in Sec.~\ref{sec:results}. 
Finally, we conclude by summarizing our progress and briefly considering
future steps.

\section{Methodology\label{sec:methodology}}
\subsection{Non-compact QED and compact QED}
There are two methods for including the electromagnetic field in the lattice simulations. 
Non-compact QED employs an electromagnetic gauge potential $A_\mu$ whose value ranges from 
$-\infty$ to $\infty$. The QED action is calculated from the gauge
potential similarly to the way it is in the continuum. 
\begin{eqnarray}
S_{\rm QED}=\frac{1}{4}\sum_{x,\mu,\nu}(\partial_\mu A_\nu(x)-\partial_\nu A_\mu(x))^2 \ ,
\end{eqnarray}
but the derivative is replaced by a finite-difference approximation.
The advantage of this method is that in quenched-QED calculations, 
the $U(1)$ configurations
are generated independently without using a Markov chain. 
One can generate $A_\mu$ in momentum space from a
Gaussian distribution and Fourier transform the results back to coordinate space.
Therefore, there are no autocorrelations among the $U(1)$ configurations. 
However, this is not true in dynamical QCD+QED simulations,
although there are some techniques to decrease the autocorrelations between
consecutive $U(1)$ configurations~\cite{Borsanyi:2014jba}. 
In addition, the non-compact $U(1)$ formalism requires gauge fixing 
during the $U(1)$ gauge-field generation process.

The other method, the compact QED formalism, uses a
complex number to represent the lattice $U(1)$ field.  The QED action is written as
\begin{eqnarray}
S_{\rm QED}&=&\beta\sum_{x,\mu,\nu}(1-\Box_{\mu\nu}) \ , 
\end{eqnarray}
where the $\beta=1/{e^2}$ and $\Box_{\mu\nu}$ is the $U(1)$ plaquette. 
The compact-QED action is quite similar to the Wilson QCD action, 
which allows us to reuse the preexisting $SU(3)$ code
with minimum change (but loss of efficiency). 
In addition, the $U(1)$ gauge fixing can be done after configurations are generated, 
which makes it somewhat easier to implement the dynamical QCD+QED algorithm. 
Because of these advantages of the compact QED, we implemented our 
dynamical QCD+QED algorithm with the compact QED formalism. 
However, it should be kept in mind that the compact $U(1)$ formalism has 
photon self-interactions as a lattice artifact, and we must 
treat this artifact carefully in our analysis. 

\subsection{Dynamical QCD+QED simulation algorithm}
We start this section with a brief recap of the RHMC algorithm for dynamical 
QCD simulations.  We 
then extend the algorithm to dynamical QCD+QED.  We follow the notation in 
Ref.~\cite{Bazavov:2009bb}.  In continuum QCD, the expectation value of an observable 
$\hat{O}$ is given by a path integral:
\begin{align}
\langle{\hat{O}}\rangle&=\frac{1}{Z(\beta)}\int \prod_{x,\mu} dU_\mu(x) \hat{O} {(\det M_F)}^n \exp\{-S_G\} \ ,
\end{align}
where the fermion fields are integrated out resulting
in a fermion determinant $\det M_F$ (raised to an appropriate power) and $S_G$ is the
gluon-field action. 
We generate the $SU(3)$ gluon field $U_\mu$ with probability distribution $P_{U}$ 
using effective action $S_{\rm eff}$:
\begin{align}
P_{U}&=\frac{1}{Z(\beta)}[\det M_F(U)]^n \exp \{-S_G(U)\}=\frac{1}{Z}\exp\{-S_{\rm eff}(U)\}  \ , \\
S_{\rm eff}&=S_G(U)+n {\rm Tr} \ln {M_F(U)} \ .
\end{align}
Using hybrid molecular dynamics,
one adds a conjugate momentum dependent term to $S_{\rm eff}$ to form an 
effective Hamiltonian~\cite{Gottlieb:1987mq}
\begin{align}
H(p, U)&= \sum_{x,\mu}\frac{1}{2}{\rm Tr} H_\mu^2+S_{\rm eff}(U)  \ ,
\end{align}
where $H_\mu$ is a traceless Hermitian matrix. The evolution of the system is given 
by Hamilton's equations, 
\begin{align}
\left \{ 
\begin{array}{c c l}
\dot{U}_\mu & = & iH_\mu U_\mu  \\
\dot{H}_\mu & = & iU_\mu\frac{\partial{S_{\rm eff}}}{\partial{U_\mu} } \Big |_{\rm TH} \ ,
\end{array}
\right .
\end{align}
where TH means a traceless Hermitian projection. 
The ${\rm Tr} \ln {M_F(U)}$ is handled by introducing a pseudo-fermion field $\Phi$,
but now we must deal with the inverse of ${M_F(U)}$.
\begin{align}
S_{\rm eff}&=S_G(U)+\Phi^+ M_F^{-1} \Phi \ ,
\end{align}
and the force becomes
\begin{align}
\frac{\partial{S_{\rm eff}}}{\partial{U_\mu}}&={\frac{\partial{S_G}}{\partial{U_\mu}}}-
 {\Phi^+ M_F^{-1}(U) \frac{\partial{M_F(U)}}{\partial{U_\mu}}M_F^{-1}(U)\Phi} \Big |_{\rm TH}\ .
\end{align}
The two terms on the RHS are the gauge force and fermion force, respectively. 
Both are needed to update the conjugate momentum.

To incorporate QED into the dynamical QCD simulation and obtain a 
dynamical QCD+QED algorithm, we need to implement several changes. 
We add $U(1)$ field links ($U_\mu^{\rm QED}$) and corresponding 
conjugate-momentum variables.  
We then add a QED gauge contribution to the action, 
\begin{align}
S=S^{\rm QCD}_G+S^{\rm QED}_G+\Phi^+ M_F^{-1}(U)\Phi \ , 
\end{align}
where the fermion matrix $M_F$ includes both QCD and QED effects.  
We add code to update the $U_1$ field ($U_\mu^{\rm QED}$) and 
its conjugate momentum according to Hamilton's equations. 
Finally, we change the QCD fermion force to take into account the quark
electromagnetic interactions and introduce the new QED fermion force.
In dynamical QCD+QED simulations, the sea quarks carry 
both $SU(3)$ and $U(1)$ charges. 
The QCD code we started with allowed
different staggered fermion actions including
one link, one link plus Naik term, asqtad, and HISQ.  Currently,
we combine the $SU(3)$ and $U(1)$ links and
then smear the combined link.  That is, for each charge we calculate
$U_\mu^{\E}(x)$ for that charge and then apply the desired smearing routine
to
\begin{eqnarray}
U_\mu(x)=U_\mu^{\C}(x)U_\mu^{\E}(x) \ ,
\label{eq:combined_smearing}
\end{eqnarray}
This method has been used in our quenched-QED 
simulations~\cite{Basak:2012zx, Basak:2013iw} and works well.
An alternative is to separately smear the $SU(3)$ and $U(1)$ links,
which would require some more code development.
In light of our prior success with the former method, we
stick with it for the dynamical QCD+QED simulation. 
Hamilton's equations for the
QED field and momentum are
\begin{eqnarray}
\dot{U}^\E_\mu&=&i H^\E_\mu(x) U^\E_\mu(x)  \ , \\
\dot{H}^\E_\mu&=& i U^\E_\mu(x) \frac{\partial S_{\rm eff}(U)}{\partial U^\E_\mu}  
= {\rm Tr} \left [i U_\mu(x) \frac{\partial S_{\rm eff}(U)}{\partial U_\mu} \right ] \ ,
\end{eqnarray}
where the trace is over $SU(3)$ indices.
Comparing with Hamilton's equations of $\dot{H}^\C_\mu$, we take 
the projection of trace of the force term. 
In addition, the second equal sign in the second equation comes from both 
the chain rule and Eq.~(\ref{eq:combined_smearing}), which is 
another reason why we use Eq.~(\ref{eq:combined_smearing}). 
Allowing independent smearings for QCD and QED gauge fields would require
additional code development.

\section{Results\label{sec:results}}
\begin{figure}[ht]
\centering
\includegraphics[scale=0.8]{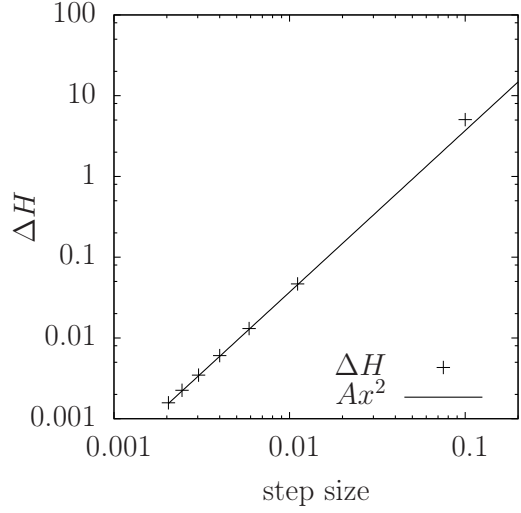}
\caption{Test of conservation of the Hamiltonian with the leapfrog integrator 
in the dynamical QCD+QED code. 
The line is proportional to the square of the step size, showing that
$\Delta H$ has the expected behavior.
The slope of the line indicates that $\Delta H$ is proportional to the square 
of the step size.}
\label{fig:int.error}
\end{figure}

We implemented the new features required
for dynamical QCD+QED simulation algorithm starting from the MILC dynamical-QCD 
simulation code. To check the new code, we studied how well the effective 
Hamiltonian is conserved. In the MILC code, we can choose different integrators 
such as leapfrog, Omelyan, etc.  We first tested the leapfrog integrator.  
On a $6^4$ grid, starting from fixed random $SU(3)$ and $U(1)$ fields, 
we ran the leapfrog algorithm with varying step sizes, but fixed unit total
trajectory length.
Figure~\ref{fig:int.error} shows that the change in the effective Hamiltonian
$\Delta H$ is proportional to the square of the step size, as expected. 
We varied the integrator and starting configuration to verify expected
behavior in several additional cases.

\begin{figure}[ht]
\centering
\includegraphics[scale=0.8]{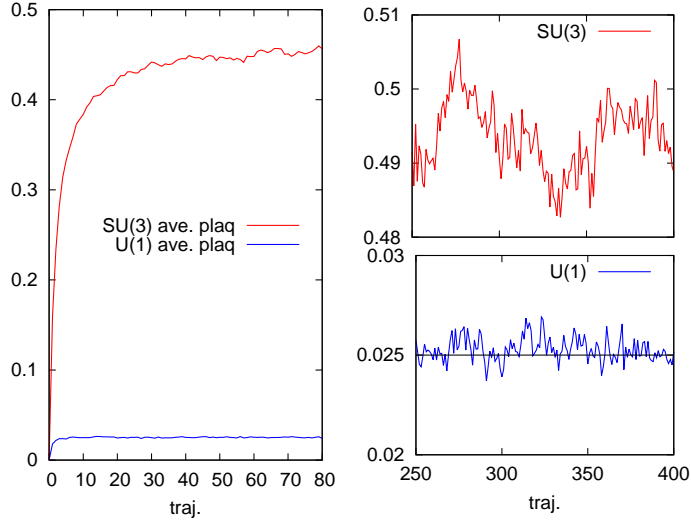}
\caption{Test of the evolution of the averaged plaquette from the dynamical QCD+QED gauge code. 
Our result is consistent with the prediction.}
\label{fig:ave.plaq.evo}
\end{figure}

After we confirmed that the integrator works correctly, we started
a test run on a $6^4$ lattice with $\beta_{\C}$=5.5 and $\beta_{\E}$=10.0. 
All of the sea quarks have the same electromagnetic charge. 
We plot the evolution of the average $SU(3)$ and $U(1)$ plaquettes 
in Fig.~\ref{fig:ave.plaq.evo}. 
Both of the QCD and QED field started from free-field configurations. 
In the left panel, we observe that the $SU(3)$ field has a different 
equilibrium time from the $U(1)$ field. The right panel shows
enlarged plots of the evolution of the $SU(3)$ (right-upper) 
and $U(1)$ (right-lower) fields. 
The black line denotes the theoretical prediction for the 
average $U(1)$ plaquette in the weak coupling limit. 
Our test shows that the evolution of the configuration is consistent with 
theoretical expectations. Similar results have been obtained in Ref.~\cite{Azcoiti:1986ry}. 

The electromagnetic contribution to the pseudoscalar-meson mass has been studied 
extensively in quenched QED.  
We generated a test ensemble with size of $12^3\times 32$. 
All three sea quarks have the same mass, 0.029 in lattice units. 
We set $\beta_{\rm QCD}=6.76$ and $\beta_{\rm QED}$ at its physical value.
We used the unimproved staggered quark action in this test.
We calculated the pseudoscalar meson mass for a few values of the
valence quark charge.  The meson, composed of a quark and anti-quark is
neutral, but its mass depends on value of the (anti-)quark charge.
We extract the electromagnetic contribution to the meson mass via: 
\begin{eqnarray}
\delta m^2 & = & m^2(e^{\rm val}\neq 0) - m^2(e^{\rm val}= 0) .
\end{eqnarray}
Chiral perturbation theory for QCD+QED predicts that to leading order
$\delta m^2$ 
is proportional to $\alpha_{\rm EM}^{\rm val}$~\cite{Bijnens:2006mk}. 
Our calculation is consistent with this.

\begin{figure}[ht]
\centering
\includegraphics[scale=0.8]{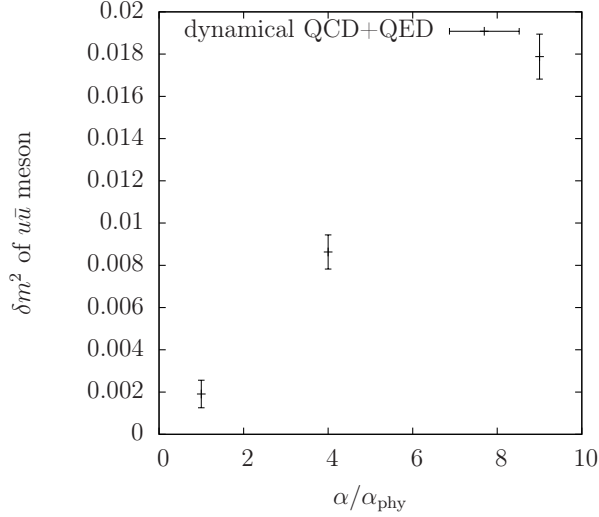}
\caption{The $\delta m^2$ calculated from different valance electromagnetic charges. 
The result is proportional to $\alpha_{\rm EM}^{\rm val}$ as expectated
from chiral perturbation theory.}
\label{fig:delta.m2}
\end{figure}
 
\section{Summary and outlook\label{sec:summary}}
In this work, we discuss the RHMC algorithm for dynamical QCD+QED 
calculations with compact QED. We implemented the algorithm based 
upon MILC dynamical-QCD code. We tested the correctness and performance 
of our code by checking the discretization error of the integrator, the 
evolution of the plaquette, and the electromagnetic contribution to 
the pseudoscalar-meson mass. 
These tests show the expected behavior and give us some confidence to
generate additional QCD+QED ensembles with larger lattice 
sizes and multiple sea-quark mass and charge combinations. 
We have also recently begun comparing our results with an independent 
coding effort of James Osborn based on FUEL.
Eventually, this will allow us to perform high-precision calculations that 
include electromagnetic effects.

\ \\
{\bf Acknowledgments:} R.~Z. thanks Thomas Blum for helpful discussions
about the dynamical QCD + QED algorithm when he was a student at
the University of Connecticut.  
We thank Claude Bernard, Ruth Van de Water and Yuzhi Liu for thoughtful 
comments on the manuscript.
R.~Z. was partially supported by NSF
Grant PHY--1212389 and DOE Grant FG02--91ER 40661.
Fermilab is operated by Fermi Research Alliance, LLC, under Contract No. DEAC02-
07CH11359 with the U.S. Department of Energy.
S.~G. was supported by DOE Grants FG02--91ER 40661 and DE-SC0010120.

\end{document}